\documentclass[sigconf]{acmart}

\AtBeginDocument{%
  \providecommand\BibTeX{{%
    \normalfont B\kern-0.5em{\scshape i\kern-0.25em b}\kern-0.8em\TeX}}}

\setcopyright{acmcopyright}
\copyrightyear{2021}
\acmYear{2021}
\acmDOI{}

\acmConference[DSHealth '21]{Joint KDD 2021 Health Day and 2021 KDD Workshop on Applied Data Science for Healthcare}{Aug 14--18, 2021}{Virtual}
\acmBooktitle{DSHealth '21] Joint KDD 2021 Health Day and 2021 KDD Workshop on Applied Data Science for Healthcare, Aug 14--18, 2021, Virtual}
\acmPrice{}
\acmISBN{}

\begin{document}

\title[Assessing putative bias in prediction of anti-microbial resistance ... under explicit causal assumptions]{Assessing putative bias in prediction of anti-microbial resistance from real-world genotyping data under explicit causal assumptions}

\author{Mattia Prosperi}
\email{m.prosperi@ufl.edu}
\affiliation{%
  \institution{Department of Epidemiology, University of Florida}
  \city{Gainesville}
  \state{Florida}
  \country{USA}
  \postcode{32611}
}
\author{Simone Marini}
\email{simone.marini@ufl.edu}
\affiliation{%
  \institution{Department of Epidemiology, University of Florida}
  \city{Gainesville}
  \state{Florida}
  \country{USA}
  \postcode{32611}
}
\author{Christina Boucher}
\email{cboucher@cise.ufl.edu}
\affiliation{%
  \institution{Department of Computer Science and Information and Engineering, University of Florida}
  \city{Gainesville}
  \state{Florida}
  \country{USA}
  \postcode{32611}
}

\author{Jiang Bian}
\email{bianjiang@ufl.edu}
\affiliation{%
  \institution{Department of Health Outcomes and Biomedical Informatics, University of Florida}
  \streetaddress{}
  \city{Gainesville}
  \state{Florida}
  \country{USA}
  \postcode{32611}}

\renewcommand{\shortauthors}{Prosperi, et al.}

\begin{abstract} Whole genome sequencing (WGS) is quickly becoming the customary means for identification of antimicrobial resistance (AMR) due to its ability to obtain high resolution information about the genes and mechanisms that are causing resistance and driving pathogen mobility. By contrast, traditional phenotypic (antibiogram) testing cannot easily elucidate such information. Yet development of AMR prediction tools from genotype-phenotype data can be biased, since sampling is non-randomized. Sample provenience, period of collection, and species representation can confound the association of genetic traits with AMR. Thus, prediction models can perform poorly on new data with sampling distribution shifts. In this work -- under an explicit set of causal assumptions -- we evaluate the effectiveness of propensity-based rebalancing and confounding adjustment on AMR prediction using genotype-phenotype AMR data from the Pathosystems Resource Integration Center (PATRIC). We select bacterial genotypes (encoded as $k$-mer signatures, i.e. DNA fragments of length $k$), country, year, species, and AMR phenotypes for the tetracycline drug class, preparing test data with recent genomes coming from a single country. We test boosted logistic regression (BLR) and random forests (RF) with/without bias-handling. On 10,936 instances, we find evidence of species, location and year imbalance with respect to the AMR phenotype. The crude versus bias-adjusted change in effect of genetic signatures on AMR varies but only moderately (selecting the top 20,000 out of 40+ million $k$-mers). The area under the receiver operating characteristic (AUROC) of the RF (0.95) is comparable to that of BLR (0.94) on both out-of-bag samples from bootstrap and the external test (n=1,085), where AUROCs do not decrease. We observe a 1\%-5\% gain in AUROC with bias-handling compared to the sole use of genetic signatures. In conclusion, we recommend using causally-informed prediction methods for modelling real-world AMR data; however, traditional adjustment or propensity-based methods may not provide advantage in all use cases and further methodological development should be sought.
\end{abstract}




\maketitle

\section{Introduction}
 Identification of antimicrobial resistance (AMR) from whole genome sequencing (WGS) data is becoming increasingly prevalent due to its ability to render high resolution about the genes and mechanisms that drive resistance and mobility; such resolution cannot be achieved using 
 traditional in vitro antibiotic phenotypic susceptibility testing (antibiogram) \cite{amr1,amr2}.  In addition, the development of miniaturized sequencing technology (Oxford Nanopore technology) will drive the emergence of WGS as the de facto method for prediction of AMR genes and resistance phenotype.  After the generation of sequence data, computational methods are used to predict the AMR genes and elements.  Several of these methods identify the presence of AMR genes using curated databases; for instance, AMRPlusPlus and the Resistance Gene Identifier align sequence data to the MEGARes and the Comprehensive Antibiotic Resistance Database databases, respectively \cite{megares,card}. Other methods predict the AMR phenotype by learning from datasets where bacteria cultured for in vitro antibiotic susceptibility testing are also sequenced, obtaining genotype-phenotype pairs; for instance, interpretable models for multiple antibiotics have been presented \citep{drouin19}, using data from the Pathosystems Resource Integration Center (PATRIC) \cite{patric}.
 
 Development of AMR prediction tools based on genotype-phenotype learning can be biased because the data repositories are made upon non-randomized sampling. Sample provenience and period of collection can affect species representation and confound the association of genetic traits with AMR \cite{oliva19}. In general, any real-world data that is observational in nature can be contaminated with different biases, since the knowledge on the data generation process or the relations among variables involved in such a process is limited \cite{prosperi20}. Learning algorithms estimate conditional probability $P(y|X)$ in various forms to predict $y$ from $X$ but the effect of variables in $X$ on to $y$ could be due to unmeasured components $Z$, which are the true causes of both $X$ and $y$. Learning models in the presence of bias usually does not affect prediction performance, i.e., $P(y|X)$ can be approximated with negligible error, regardless $X$ being the true causes of $y$. As a result, prediction models can have high performance in cross-validation, yet perform poorly in real-world settings because of sampling distribution shift, e.g. species prevalence \cite{lawrence09}.

 In this work --under explicit causal assumptions-- we evaluate the effectiveness of propensity-based rebalancing and confounding adjustment on AMR prediction with external data characterized by species, time, and location distribution shifts.

\section{Methods}
\subsection{Potential outcomes and propensity scores}
We use Rubin's notation of potential outcomes for calculating exposure or treatment effects \cite{rubin74}. We assume a sample population of $N$ individuals, who were exposed or received a treatment $T$ (binary, for ease) and an health outcome $Y$ was observed after treatment. Each subject $i$ is represented by the tuple \{$T_i,Y_i,X_i$\}, where $X$ is a vector of preexposure variables (e.g., sex, insurance status). We define $Y_{i}^0$ and $Y_{i}^1$ as the potential outcomes for person $i$ under treatment/exposures $T_i=0$ and $T_i=1$, respectively. The individual treatment effect (ITE) $\tau(x)$ is the difference in the average potential outcome for $i$ under both $T$, conditional on the preexposure vector $X=x$, i.e. $\tau(x)=\mathbb{E}[Y_{i}^1 - Y_{i}^0 \mid X_i = x]$. People cannot have two different exposures (e.g. taking/not taking a therapy or having/not having a genetic trait) at the same time, so only one --factual-- outcome can be observed.  Thus, the ITE $\tau(x)$ usually cannot be calculated as one outcome is missing. However, if one assumes that potential outcomes are independent of the exposure conditional on preexposure vector, i.e., \{$Y_{i}^1Y_{i}^0$\} $\perp T | X$, then the ITE can then be calculated under the strongly ignorable treatment assignment (SITA) assumption, as $\tau(x)=\mathbb{E}[Y^1 \mid T=1, X = x] - \mathbb{E}[Y^0 \mid T=0, X = x] 
= \mathbb{E}[Y \mid T=1, X = x] - \mathbb{E}[Y \mid T=0, X = x]$.
By assuming SITA and considering the distribution of  $X$, we can calculate the average treatment effect (ATE) $\tau_{01}$ as
$\tau_{01}=\mathbb{E}[\tau(X)]=\mathbb{E}[Y\mid T=1]-\mathbb{E}[Y\mid T=0]$. The ATE can also be conveniently expressed as an odds ratio, however, the marginal and conditional effects are not guaranteed to coincide, i.e., the treatment effect measure is not collapsible \cite{didelez21}. ITE and ATE can be calculated with $x$ being equally matched in exposure/control groups but stratification becomes infeasible as the number of dimensions increases. One way to match individuals without crisp stratification is via propensity scores. The propensity score $\pi(x)$ represents the probability of receiving a treatment or being exposed to $T=1$ (assuming that the alternative is no treatment or exposure $T=0$) conditioned on the preexposure covariates $X$, and is denoted as $\pi(x) = P(T = 1|X = x)$.
Through the conditional probability $\pi(x)$, we can balance the probability of being exposed either to $T=1$ or $T=0$ given $X=x$. The propensity score can be estimated using any regression technique from logistic regression to Bayesian additive regression trees \cite{hill11}. The original dataset can be then rebalanced with respect to the exposure by matching pairs of exposed/unexposed subjects with similar propensity scores, i.e., performing propensity score matching (PSM). Several algorithms can be used for PSM, such as nearest neighbor or caliper \cite{austin14}. 

\subsection{Proposed approach}
\begin{figure}
  \centering
  \includegraphics[width=0.8\linewidth]{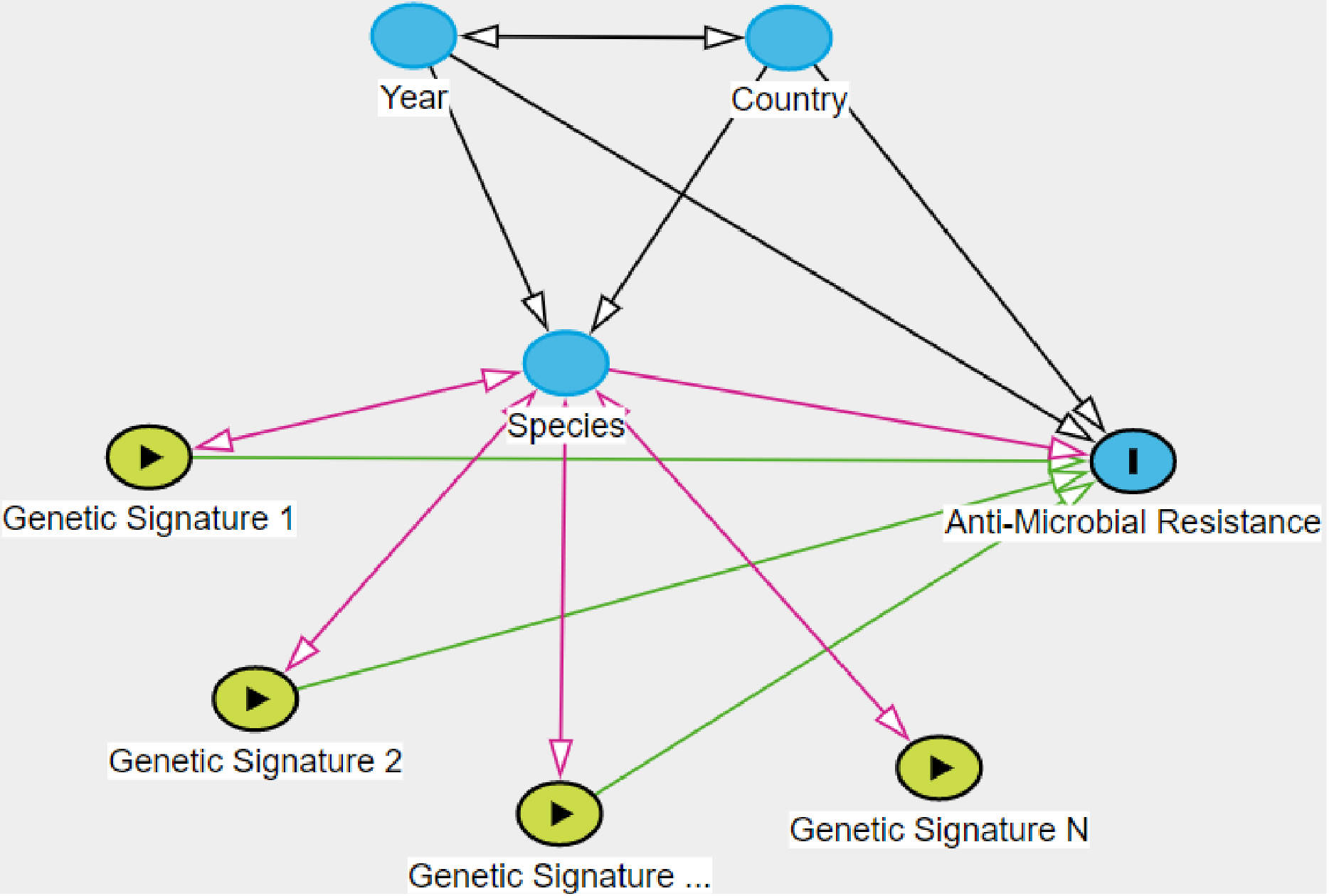}
  \caption{Causal structure (partially directed acyclic graph) for the effect of genetic signatures of bacteria on to anti-microbial resistance, confounded by country of sampling, year, and species.}
  \label{fig1}
  \Description{}
\end{figure}

As previously described, the effect of a genetic trait of a bacterium on the odds of carrying AMR could be confounded by the fact that the genetic trait itself is characteristic of a species that has been unevenly sampled among antimicrobial-resistant cases. The species could be found to be more prevalent in specific locations or periods of time, and these two variables could affect the overall prevalence of AMR in a population or environment. By considering AMR as an outcome, we can model the genetic trait akin to an exposure confounded by species, locations and times. Figure \ref{fig1} shows the causal relationships among the aforementioned variables using a partially directed acyclic graph (PDAG). By using d-calculus, e.g. the back-door criterion \cite{pearl16}, it is possible to identify a minimal set of variable adjustments for the total effect \cite{percovic15} of one or more genetic signatures on to AMR. Depending on the directions of the arrows between species and genes, the minimum adjustment set would include only the species, or the species, country and year together. Additional arrows between genetic signature pairs or between genetic signatures and the AMR phenotype do not change this adjustment set. Therefore, it is safe to use the species, country and year together as adjustment covariates for effect estimation since they do not add bias for (counterfactual) prediction.

Using this explicit causal structure, we define the following data rebalancing and adjusting strategies to handle bias: (i) simple covariate adjustment pooling together all $k$-mers with country $c$, year $y$, and species $s$ variables; (ii) data rebalancing using PSM on $c,s,y$ with respect to AMR phenotype $r$, i.e. $P(r|c,y,s)$; (iii) covariate ranking and pre-selection based on calculation of ATE of each $k$-mer adjusted by $c,s,y$ with respect to AMR; (iv) data rebalancing using $k$-mer clustering (since $k$-mers are differently distributed within species, this would balance species oversampling regardless the contribution of single $k$-mers to AMR). For the PSM, we compare three different techniques: nearest neighbor matching with either logistic regression (LR) or random forest (RF) score learning, and coarsened exact matching. The $k$-mer clustering is done via logistic principal component analysis (LPCA) \cite{landgraf2020dimensionality} of the $k$-mer vectors, then passed on to an agglomerative hierarchical average linkage algorithm, selecting the best number of clusters by maximizing the average silhouette values. Each bias-handling strategy can be applied in conjunction with any machine learning model of choice (whereas the base model is the one that uses only $k$-mers as input to predict AMR), and here we use boosted LR (BLR) and RF \cite{friedman00,breiman01}.

\begin{table}
\footnotesize
  \caption{Sampling characteristics of the PATRIC genotypes with known tetracycline antibiogram results (n=10,936)}
  \label{tab1}
  \begin{tabular}{|p{1cm}|p{1.4cm}|p{1.7cm}|p{1.2cm}|p{1.2cm}|}
    \toprule
    Set (N) & Domain & Variable & Res N (\%) & Sus N (\%) \\
    \midrule
    Training & Country & Canada & 131 (59\%) & 90 (41\%) \\
    (9,851) & & Israel & 70 (18\%) & 318 (82\%) \\
    & & South Africa & 281 (56\%) & 224 (44\%) \\
    & & UK  & 567 (43\%) & 750 (57\%) \\
    & & USA & 2150 (51\%) & 2059 (49\%) \\
    & & Other-Unkn. & 1278 (40\%) & 1933 (60\%) \\
    & Year & 1905-2006  & 981 (70\%) & 492 (30\%) \\
    &  & 2007-2010  &  606 (56\%) &  472 (44\%) \\
    &  & 2011-2014  &  1363 (48\%) &  1451 (52\%) \\
    &  & 2015-   &  1479 (34\%) &  2877 (66\%) \\
   & Species & Streptococcus  &  919 (42\%) & 1254 (58\%) \\
    & & Salmonella & 1102 (54\%) & 921 (46\%) \\
    & & Staphylococcus & 322 (16\%) & 1693 (84\%) \\
    & & Other & 2134 (67\%) & 1056 (33\%) \\
    \midrule
    Test & Country/Year & USA, 2015- & 702 (65\%) & 383 (35\%) \\
    (1,085) & Species & Streptococcus  & 1 (100\%) & 0 (\%) \\
    &  & Salmonella  & 4 (19\%) & 17 (81\%) \\
    &  & Staphylococcus  & 0 (0\%) & 5 (100\%) \\
    & & Other  &  698 (67\%) & 350 (33\%) \\
  \bottomrule
\end{tabular}
\end{table}

\subsection{Data and experimental settings}
We extract genotype-phenotype AMR data and metadata (country, year, species) from the PATRIC repository (\url{https://www.patricbrc.org}). For this work, we use the tetracycline class of antibiotics as a use case (representing a whole antibiotic class as opposed to single drug compounds, and among the most frequent in the data base). Only bacterial genomes with an associated laboratory test are considered, and only entries that have at least a known metadata attribute. The bacterial genomes (nucleotides) are encoded into $k$-mers (strings of fixed length $k$), with $k=17$, chosen on the basis of optimization from prior studies. Since the total number of $k$-mers can be overwhelming (hundreds or tens of millions depending on the species diversity), and since $k$-mers have very long-tailed distributions of frequencies, we select the top 10,000 $k$-mers based on frequency, and the top 10,000 $k$-mers based on information gain with respect to the AMR phenotype (sole training set). These 20,000 are then re-ranked according to each rebalancing strategy, e.g. calculating odds ratio on the basis of PSM for strategy (i) or using the adjusted ATE for strategy (iii), and then the top 5,000 are fed to the internal feature selection of BLR or RF. Countries with low frequency (<90\% cumulative, after ordering) are grouped into a single 'other' category. Years are divided into quartiles, and species are grouped at the genus level. A test set is created by selecting instances from a single country, collected in the most recent quartile year. This procedure respects the overlap condition for PSM estimation and yet induces a shifted dataset.

We assess discriminating ability of models using the area under the receiver operating characteristics (AUROC), estimated robustly using out-of-bag samples (10 bootstrap replicates) and on the external shifted dataset. All parameter optimizations (number of boosting iterations and of random trees, between 100 and 6,000) and rebalancing procedures are done internally to the bootstrapping. The PATRIC data preprocessing, LPCA, $k$-mer counting and extraction --made using KMC3 \cite{kokot17}-- are executed on the University of Florida's HiPerGator3 high-performance computing cluster (\url{https://www.rc.ufl.edu/services/hipergator/}). All subsequent analyses are performed on an Intel laptop with i9-10885H CPU at 2.4GHz and 32MB RAM, using R \cite{r}, including libraries cluster, mboost, MatchIt, parallelDist, ranger, ROCR. R scripts and datasets are available at \url{https://github.com/DataIntellSystLab/amr_bias}.

\begin{figure}[b]
  \centering
  \includegraphics[width=0.6\linewidth]{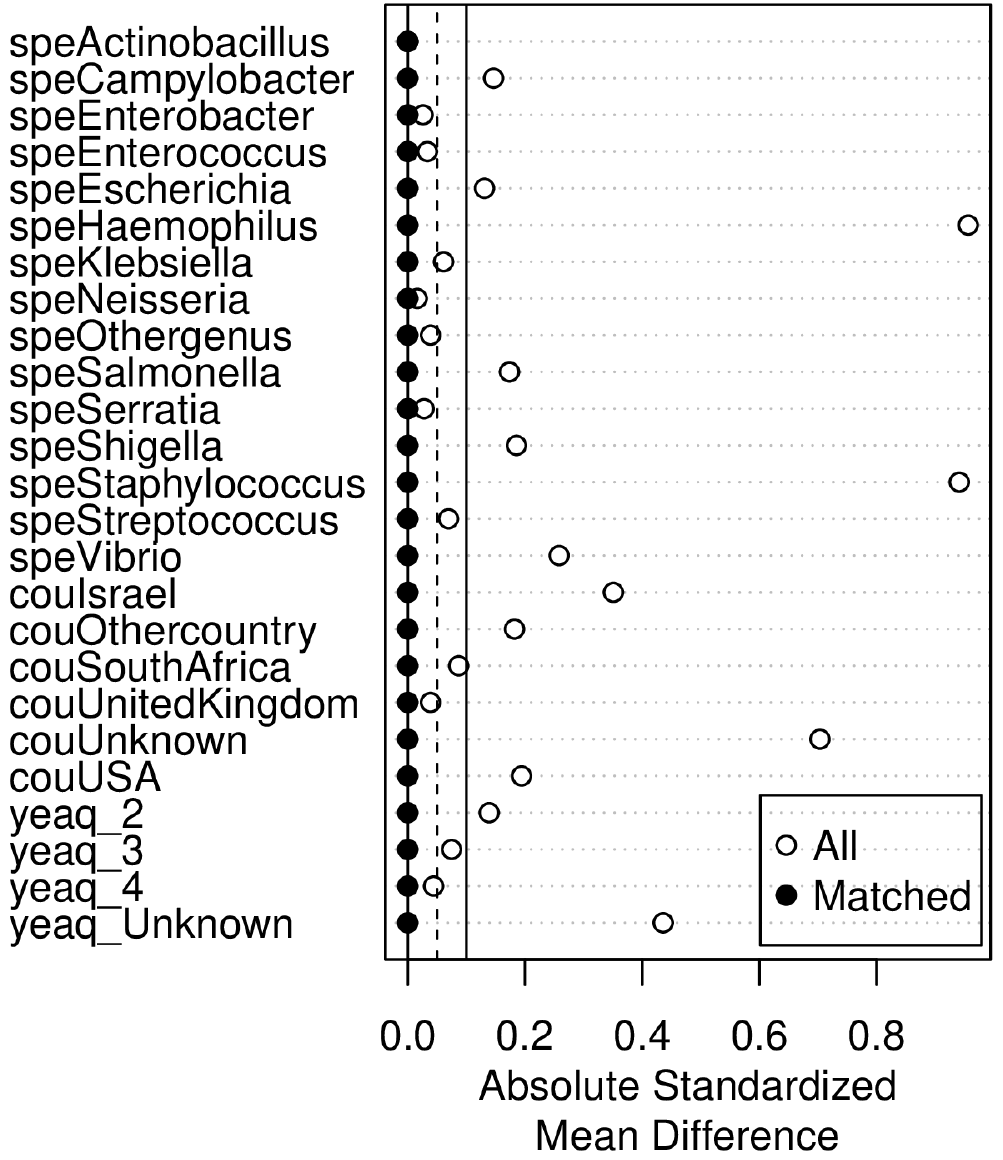}
  \caption{Propensity score matching of sampling variables (spe=species' genus, cou=country, yeaq=quartile year) with respect to AMR phenotype. Black dots show absolute standardized mean difference for the rebalanced data, whilst white dots show the original data.}
  \label{fig2}
  \Description{}
\end{figure}

\begin{figure}
  \centering
  \includegraphics[width=0.8\linewidth]{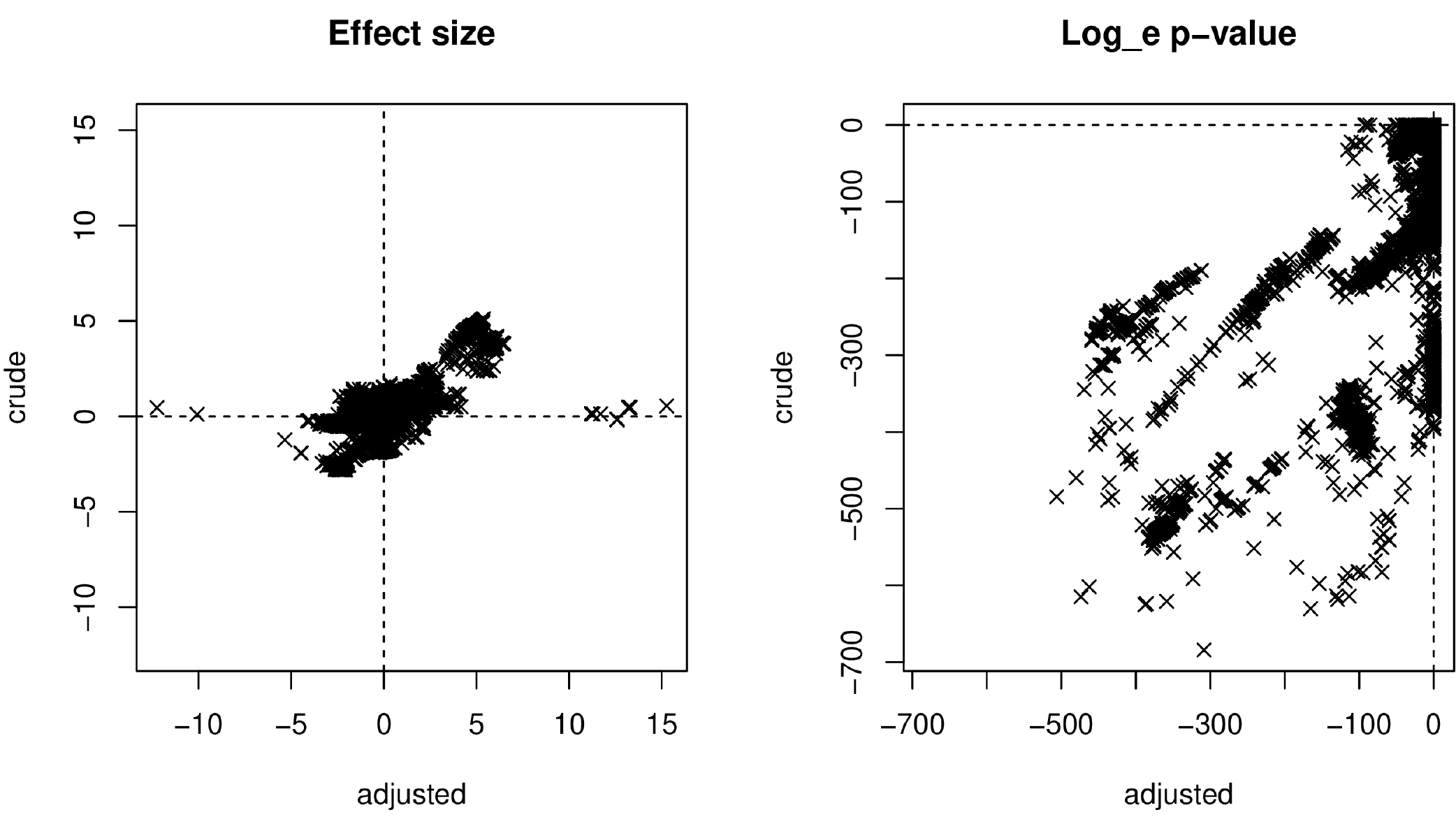}
  \caption{Adjusted vs. crude ATE estimation of each genetic signature (n=20,000) with respect to the AMR phenotype. Left panel shows scatterplot of effect sizes (dashed lines cross zero effect), whilst the right panel compares p-values (dashed lines are p=0.05).}
  \label{fig3}
  \Description{}
\end{figure}

\section{Results}
After querying PATRIC (currently hosting over 80,000 bacterial genomes) according to the inclusion criteria, we obtain 10,936 genotypes and metadata pairs with an antibiogram test result for tetracycline. Of note, additional 8,815 samples meet the inclusion criteria but do not have a legit AMR phenotype. After removing the samples from USA collected in the most recent year quartile, the training set comprises 9,851 instances, while the external test set includes 1,085. Table~\ref{tab1} shows sampling characteristics of the training and test set, stratified by AMR phenotypes; it is evident how species, country and year differ sensibly among the datasets and strata.

After performing PSM on the AMR phenotypes (with coarsened exact matching being the best method compared to LR or RF nearest neighbor), the sampling variables seem to be perfectly rebalanced with respect to the phenotype (Fig~\ref{fig2}). The ATE estimation performed on each $k$-mer shows that there is dispersion between the crude and the bias-adjusted values, but the direction of the effects does not seem to be strongly affected, with only 8.1\% of ATEs changing direction when adjusted (Fig~\ref{fig3}). Of note, 20.2\% of the $k$-mers have a p-value>0.05 when the crude p-value<=0.05.

\begin{figure}[b]
  \centering
  \includegraphics[width=0.5\linewidth]{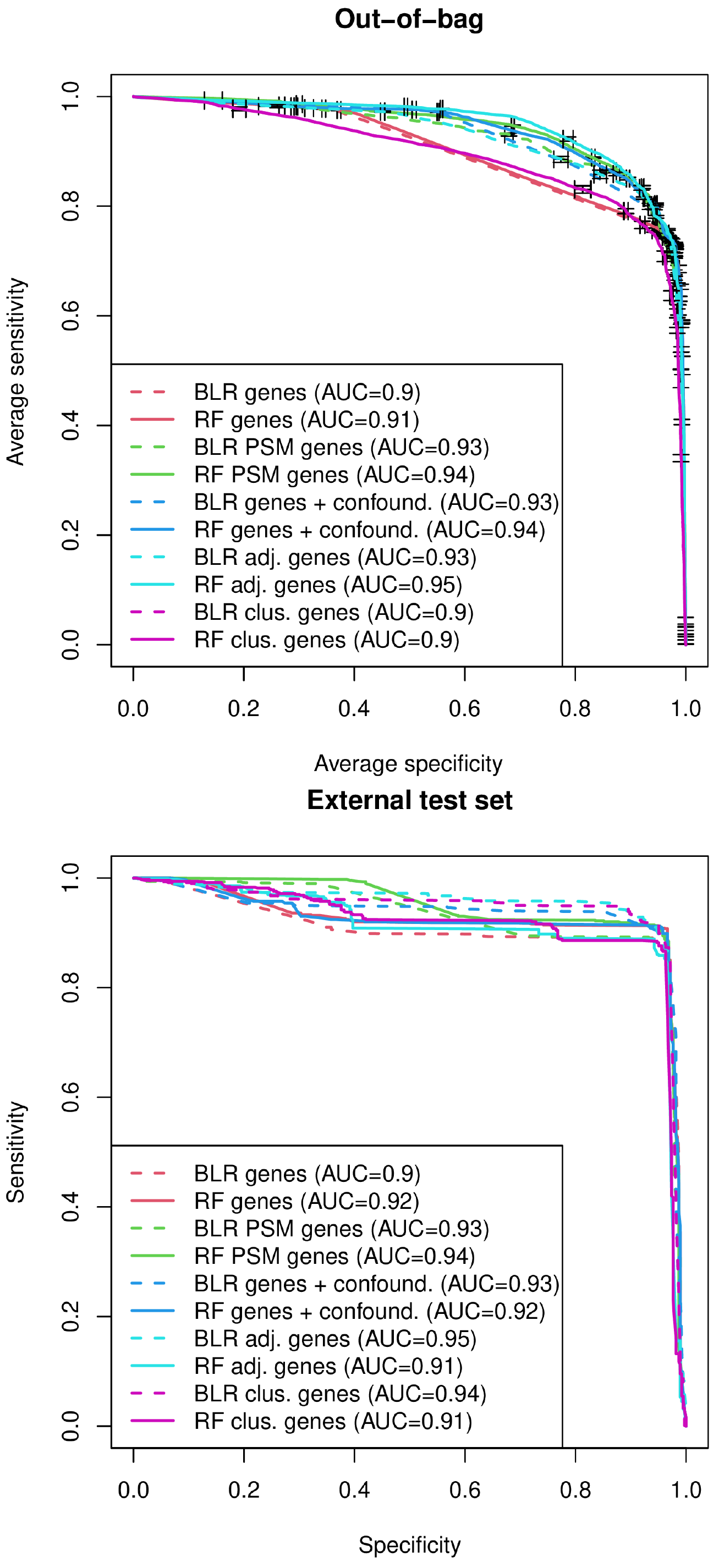}
  \caption{Discriminating ability in prediction of AMR phenotype by random forest and boosted logistic regression with different bias-handling strategies (vs. none) on ten out-of-bag samples (top panel) and external test set (bottom panel).}
  \label{fig4}
  \Description{}
\end{figure}

The discriminative performance of prediction models on the out-of-bag samples drawn from the training data shows that RF is comparable to BLR using both bias-handling strategies and no bias-handling. The optimal number of boosting iterations for BLR and number of trees for RF stabilizes around 3,000. For BLR, AUROCs range from 0.93 obtained using any of the bias-handling strategies (i)-(iii) to 0.9 without bias handling. For RF, AUROCs range from 0.95 with strategy (iii) to 0.9 without bias handling. The bias-handling strategies yielded an increase in AUROC of 3\% to 5\% in the out-of-bag validation. When applying the models on the external test dataset, AUROC did not decrease significantly. The gain of the bias-handling strategies was 1\%-5\%. Fig~\ref{fig4} shows all AUROCs for BLR and RF with respect to each strategy.

\section{Discussion}
In this work, we explore different strategies to assess putative bias in prediction of AMR from real-world data under explicit causal assumptions. In the PATRIC genotype-phenotype data relative to tetracycline antibiotics, we find strong evidence of sampling and confounding bias with respect to species, country, and year of collection. Nonetheless, the change in effect for each individual genetic signature with respect to ATE when comparing crude and adjusted estimation is mild. Genetic signatures likely already contain information that defines the species, but they likely cannot account for prevalence changes in AMR by country and year.
Causally-informed methods improve discriminating ability of AMR as compared to approaches that use only genetic signatures, but the gain is 5\% or less for most configurations.

One limitation of this approach is that we analyze only 20,000 gene signatures out of over 40 million. Half of the pre-selection is based on information gain, which is basically a crude ATE and could be biased. Another possible issue is that the test set is a single realization and thus the AMR distribution shift is non-generalized.

Regardless the use of bias-handling, nonlinear ensemble classifiers like the RF can be computationally burdensome to run even after training. Given that bacterial sequencing is being increasingly done in real-time and mobile settings \cite{oliva20}, development of fast classifiers is warranted, thus the choice of BLR seems favorable.

Here, we restrict interest to WGS but extension to metagenomics data warrants future investigations. Studying bias in this setting is complicated by the prevalence of multiple -- sometimes putative -- AMR pathogens and species, which will confound the $k$-mer composition. Moreover, geographical location has shown to have a substantial effect of metagenomes \cite{PASOLLI2019649,doi:10.1080/19490976.2020.1752605} so we hypothesize that our results will be more pronounced in a metagenomics setting.

In conclusion, we recommend the usage of causally-informed methods for the development of computational AMR prediction models using observational, real-world data, when there is evidence of bias in species sampling; however, traditional adjustment or propensity-based methods may not provide advantage in all use cases and further methodological development should be sought.

\begin{acks}
This work was in part supported by US grants NIH NIAID R01AI145552, NIH NIAID R01AI141810, and NSF SCH 2013998.
\end{acks}

\bibliographystyle{ACM-Reference-Format}
\bibliography{camera_ready}


\end{document}